\documentclass{article}
\usepackage{graphicx}
\usepackage{amssymb}
\usepackage{amsmath}
\newcommand{\dd}[1]{\ensuremath{\,{\mathrm{d}#1}}}%
\newcommand{\as}{\ensuremath{\alpha_\mathrm{s}}}
\newcommand{\epem}{\ensuremath{\mathrm{e}^+\mathrm{e}^-}}
\newcommand{\eV}{\hbox{\ensuremath{\mathrm{e\kern-0.11em V}}}}%
\newcommand{\GeV}{\hbox{\ensuremath{\mathrm{G}}\eV}}%
\newcommand{\eg}{{\it eg}}%
\begin{document}
\pagestyle{plain}
\newcount\eLiNe\eLiNe=\inputlineno\advance\eLiNe by -1
\title{QCD at LEP\footnote{Presented at XXXIII Intl. Symposium on Multiparticle
Dynamics, Cracow, Poland, 5--11 Sept., 2003. To appear in {\it Acta Physica Polonica.}}}
\author{W.J. Metzger  \\
{\small University of Nijmegen, Toernooiveld 1, 6525\ ED\ \ Nijmegen, Netherlands}}
\maketitle
 
\begin{abstract}
 Several preliminary QCD results from
  \epem\ interactions at LEP
 are reported. These include studies of
  event shape variables, which are used to
  determine $\alpha_\mathrm{s}$ and for studies of the validity of power corrections.
  Further, a study of color reconnection effects in 3-jet Z decays is reported.
\end{abstract}
 
 
\section{\boldmath{$\as$} from event shape variables}
Variables which quantify in some way the distribution of the particles of an event in momentum space,
known as shape variables, are sensitive to the amount of hard gluon emission in the event.
The distributions of such variables are therefore sensitive to the value of the strong coupling
constant, \as. The distributions expected by QCD can be fit to the data distributions in order to
measure \as.
 
To be useful for this purpose, the variables should be infrared and collinear safe and insensitive
to the electroweak physics which produces the event.
Examples of such variables are $\tau=1-T$, where $T$ is the thrust;
the scaled heavy jet mass, $M_\mathrm{H}$;
the total and wide jet broadening, $B_\mathrm{T}$ and
$B_\mathrm{W}$;
the C-parameter, $C$;
and the jet resolution parameter, $y_{23}$, which is the value of $y_\mathrm{cut}$ in the
Durham algorithm at which the event classification changes from 2-jet to 3-jet.
 
All four LEP collaborations have measured these distributions at various center-of-mass energies
and used them to determine \as.  The LEP QCD Working Group has performed a preliminary simultaneous
fit to all of these distributions \cite{lepwg}, which is described here.
Such a combined fit allows a consistent treatment of the theoretical predictions and uncertainties,
as well as of correlations between variables and energies.
 
Each experiment measures the shape variable distributions and corrects them for detector resolution and
acceptance, background, initial state radiation, {\it etc.}  The theory predictions are calculated
and, since these predictions are at parton level,  corrected for hadronization using a parton shower
Monte Carlo (MC) program such as {\sc pythia, herwig, ariadne}.
The corrected theory predictions are then
fit to the corrected experimental distributions to determine \as.
 
To ${\cal O}(\as^2)$ the distribution of shape variable $y$ is given by
\[
     f_\mathrm{pert}(y;\as)  \equiv
                     \frac{1}{\sigma_\mathrm{tot}}{\frac{\dd{\sigma}} {\dd{y}}}
              =   \frac{\as}{2\pi}{A(y)}
                      + \left(\frac{\as}{2\pi}\right)^2
                       \left[{B(y)}+2\pi\beta_0 \ln\left(\frac{\mu^2}{s}\right)
       {A(y)}\right]
\]
where $\beta_0=\frac{33-2n_\mathrm{f}}{12\pi}$, with $n_\mathrm{f}$ the number of active flavors,
and
      $\mu = {x_\mu}\sqrt{s}$ is the renormalization scale,  $x_\mu$ providing a parameter to use
to vary the scale.
Integration  of the ERT matrix elements gives the values of  $A$ and $B$.
This describes the data well in the multi-jet region, but not in the 2-jet region, which corresponds to
small values of $y$, where emission of softer gluons is important.
They can be included by summing, to all orders of \as, the
leading and next-to-leading order logarithmic terms in the expansion of
$R(y;\as)=\int_0^y f_\mathrm{pert}\dd{y}$ in terms of $L=\ln(1/y)$.
The two calculations can be combined if one is careful to avoid double counting.
This is advantageous since it allows use of a fit range extending into the 2-jet region.
However, it is not without
theoretical uncertainty.
There are two ``matching schemes'' to do this, the Log-R and the Modified Log-R schemes.
The latter forces $R$ to vanish above the kinematical maximum value of $y$
by replacing $L$ by
\[
   L^\prime =
    \frac{1}{p} \ln \left[ \left(\frac{1}{{x_\mathrm{L}}y}\right)^{p} - \left(\frac{1}{{x_\mathrm{L}}y_\mathrm{max}}\right)^{p} + 1 \right]
\]
where the parameters $p$ and $x_\mathrm{L}$ allow variation of the incorporation of the kinematical limit.
 
The data samples available cover three center-of-mass energy ranges:
 20--85\,\GeV\ from radiative Z decays where the radiated photon is removed from the event;
 $M_\mathrm{Z}$; and
 133--206\,\GeV.
At present only L3 measurements are used from the lowest energy range, although OPAL has recently
released preliminary measurements \cite{opalLow}. The values of \as\ from this new OPAL analysis agree
well with the current world average  
and are therefore not expected to have a large effect on the combination.
 
%
 
\begin{figure}[htbp]
\begin{center}
   \includegraphics[width=0.8\textwidth,bbllx=10,bblly=40,bburx=460,bbury=577,clip=]{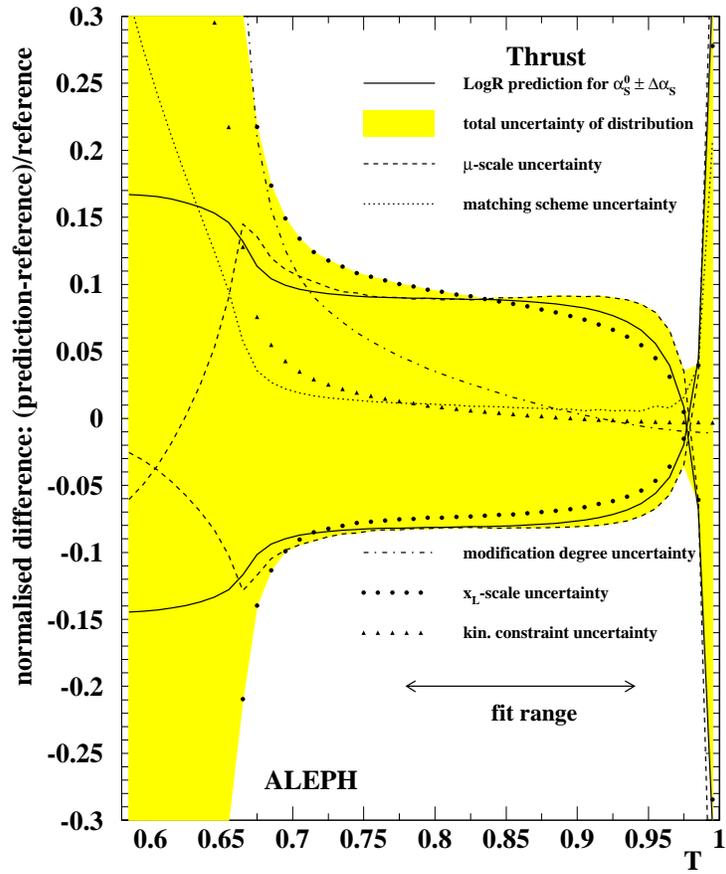}
\end{center}
\caption{Variations to determine the theoretical uncertainty on                   \as.}
\label{fig:vary}
\end{figure}

At present 6 shape variables at 14 energies are used giving 194 measurements in total, not all
variables being available for all energies/experiments. To perform the fit,
the covariance matrix ($194\times194$) of these measurements is needed.
It is composed of four contributions:
$V_{ij}=V_{ij}^\mathrm{stat}+V_{ij}^\mathrm{exp}+ V_{ij}^\mathrm{had}+V_{ij}^\mathrm{th}$.
The first term, the statistical uncertainty, is the easiest to evaluate.  It is certainly uncorrelated
between experiments and between energies.
The second term is the systematic uncertainty in the experimental measurement. It is uncorrelated between
experiments. Correlations within an experiment are taken as the ``minimal overlap'',
$V_{ij}^\mathrm{exp}=\min(V_{ii}^\mathrm{exp},V_{jj}^\mathrm{exp})$.
The third term is the systematic uncertainty in the hadronization correction of the perturbative
calculations.  While one might expect this to be correlated between experiments, since they all use the
same programs, it seems that these correlations are small. This is presumed to be due to the fact that each
experiment uses a different tuning of MC 
parameters.  These uncertainties are estimated from
a comparison of the corrections found using {\sc pythia, herwig, ariadne}. Large fluctuations are
seen in the values from energy to energy, presumably arising from statistical limitations in the
Monte Carlo samples.  Accordingly, they were smoothed assuming a $1/Q$ dependence, as suggested by
power corrections, and all correlations were assumed to be zero.
The fourth term is the uncertainty in the perturbative prediction due to neglect of higher orders.
There are certainly very large
correlations not only between experiments, but also between energies and shape variables.  However,
attempts to fit including such large correlations lead to highly unstable results.  Accordingly, all
correlations were set to zero.
To evaluate the diagonal elements, the following procedure was followed.
First, a nominal value of \as\ was chosen,
and the shape variable distributions were calculated for the default values  ${x_\mu}=x_\mathrm{L}=p=1$
using the modified Log-R scheme, giving the nominal distribution.
Distributions were then calculated varying the parameters ${x_\mu}$, $x_\mathrm{L}$ and $p$
as well as using the Log-R scheme.  The difference of these distributions
with the nominal one is shown for a typical case \cite{aleph} in Fig.~\ref{fig:vary}.
Then the value of \as\ is varied using ${x_\mu}=x_\mathrm{L}=p=1$  to determine values of \as\
which produce changes in the distribution as large as the largest differences found within the fit range.
The difference between this value of \as\ and the nominal value is  taken as the uncertainty on \as.

\begin{figure}[htbp]
\begin{center}
   \includegraphics[width=0.87\textwidth]{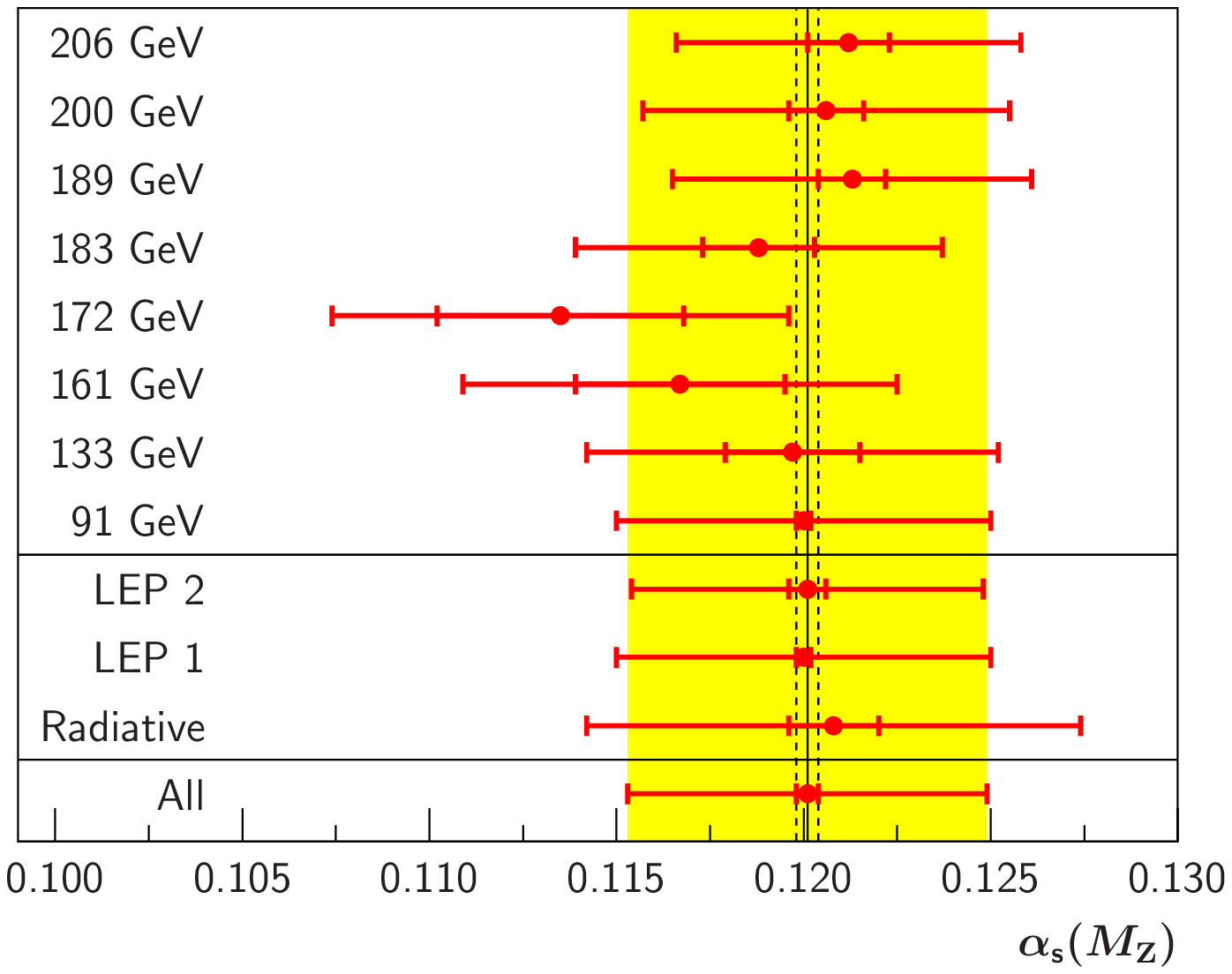}
\\ 
   \includegraphics[width=0.87\textwidth]{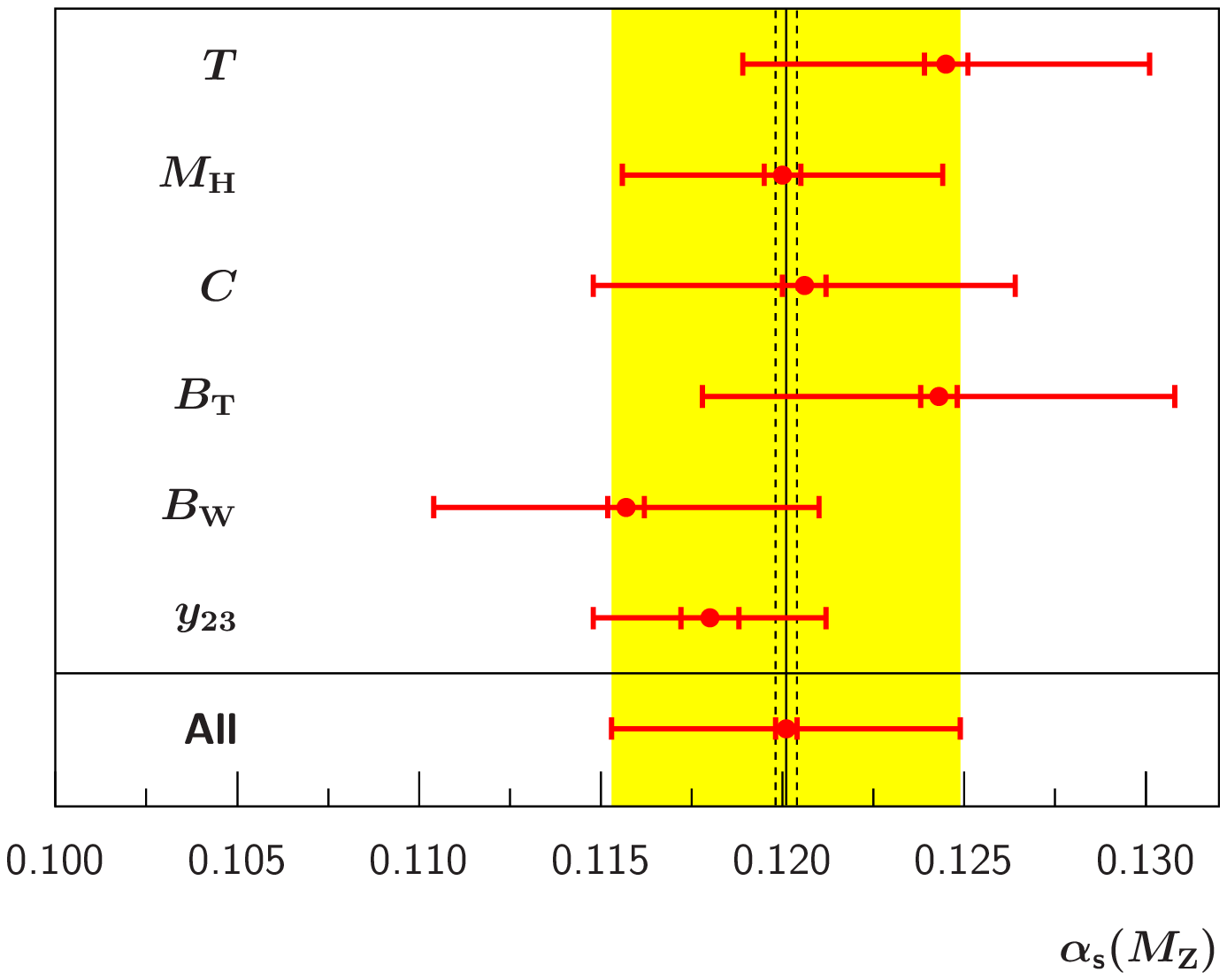}
\end{center}
\caption{Results for \as\          at different energies and using different shape variables.}
\label{fig:as}
\end{figure}

Using the above-described covariance matrix a least squares fit is performed.  Because of the assumptions
made on lack of correlations in the hadronization and theory uncertainties, the uncertainties found by
the fit can not be expected to be correct.  To determine the final hadronizaton uncertainty, the combination
is performed 3 times, once for each of the MC programs.  The result for \as\ is found using
{\sc pythia}, the uncertainty from the rms of the results.  The theory uncertainty is found by repeating
the fit twice, using for all points
$\alpha_{\mathrm{s}i}\pm\surd{V_{ii}^\mathrm{th}}$.
The theory uncertainty is taken as half the difference.
The result is
\[
   {\as(M_\mathrm{Z}) = 0.1201 \pm 0.0003\,\text{(stat)} \pm 0.0009\,\text{(exp)}
                               \pm 0.0009\,\text{(had)}  \pm 0.0047\,\text{(th)}}
\]
Note that the theory uncertainty dominates.
The results from different energies and shape variables are consistent, as shown in Fig.~\ref{fig:as}.
 
\section{Power corrections}
The power correction ansatz 
parametrizes the unknown behavior of \as\ below an infrared matching scale,
$\mu_\mathrm{I}$, by an average, $\alpha_0=\frac{1}{\mu_\mathrm{I}}\int_0^{\mu_\mathrm{I}}\as(k)\dd{k}$.
This leads to a power term, $P\propto1/\sqrt{s}$ which shifts distributions of shape variables:
$f(y) = {f_\mathrm{pert}}(y-{c_y P})$ and increases their moments:
$\langle y  \rangle  =  {\langle y   \rangle_\mathrm{pert}} + {c_y P}$ and
$\langle y^2\rangle  =  {\langle y^2 \rangle_{\text{pert}}} + {2\langle y \rangle_\mathrm{pert}} \,{c_y P}
$.
The factor $c_y$ is a known factor, different for each shape variable,
but $P$ is supposed to be universal.
 
\begin{figure}[htbp]
\begin{center}
    \includegraphics[width=0.65\textwidth]{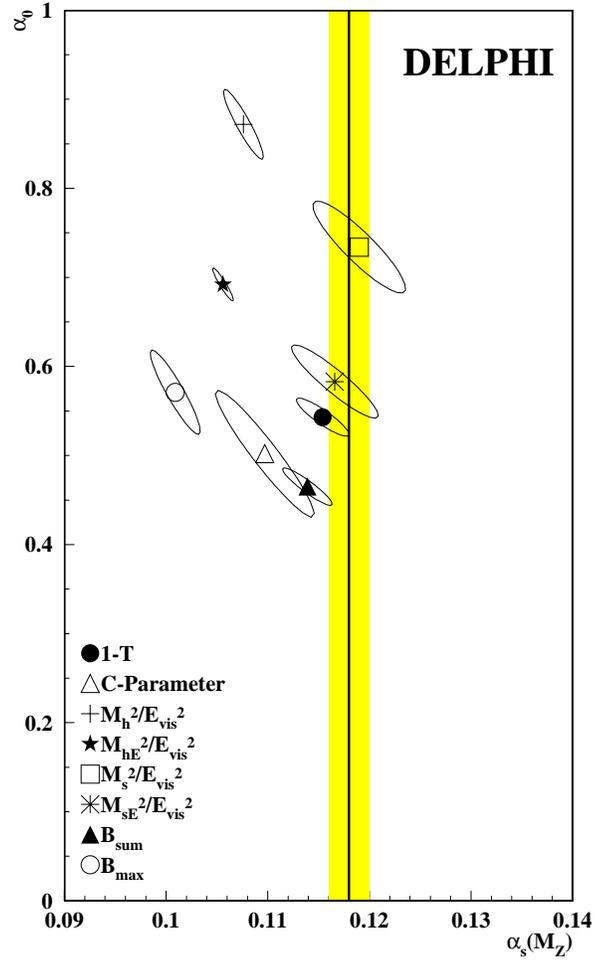}
\end{center}
\caption{Results of analysis of shape-variable distributions using a power correction ansatz.}
\label{fig:powerdist}
\end{figure}

DELPHI \cite{delphi19,delphi26} has analyzed both the distributions of a number of shape variables
and their first moments.  L3 \cite{l3qcd} has analyzed both first and second moments.
The first moments of different shape variables
result in consistent values of \as, but differences of around 20\%
are observed in the values of $\alpha_0$. The situation is much worse in the analysis of the
distributions, as shown in Fig.~\ref{fig:powerdist}, and of the second moments.
 
While one would have hoped that power corrections could have been used instead of Monte Carlo models
for the hadronization corrections of shape variables,
the results indicate that power corrections provide only
a semi-quantitative description.

\section{Color reconnection in 3-jet Z decays}
So-called rapidity gap events have been observed in ep and $\bar{\mathrm{p}}\mathrm{p}$ events and are
attributed to color-singlet exchange.  OPAL has investigated \cite{gary}
whether such effects exist in \epem\ events by studying directly
the distribution of particles within the gluon jet of 3-jet Z decays.
L3 has taken a different approach \cite{rapgap}.
 
The occurence of a color-singlet exchange within the gluon jet breaks the jet into two pieces, one
which itself is a color singlet, and another which forms a color singlet together with the quark and
antiquark.  This has an effect on the color flow between jets.  To quantify this, asymmetries are
constructed comparing the flow between q and g with that between q and $\bar{\mathrm{q}}$, \eg,
$A_{12}^\mathrm{B} =(-B_{12}+B_{23}+B_{31})/(B_{12}+B_{23}+B_{31})$, where $B_{ij}$ is the smallest
angle between two adjacent particles in the region between jets $i$ and $j$, excluding particles within
cones of $15^\circ$ about the jet axes.  A ``Mercedes'' topology is required with jet 3 identified as
the gluon jet by a b-tag for jets 1 and 2 and an anti-b-tag for jet 3,
resulting in a sample of 2668 events
with a gluon jet purity of about 78\%
 
\begin{figure}[htbp]
\begin{center}
  \hfil
  \includegraphics[width=0.72\textwidth]{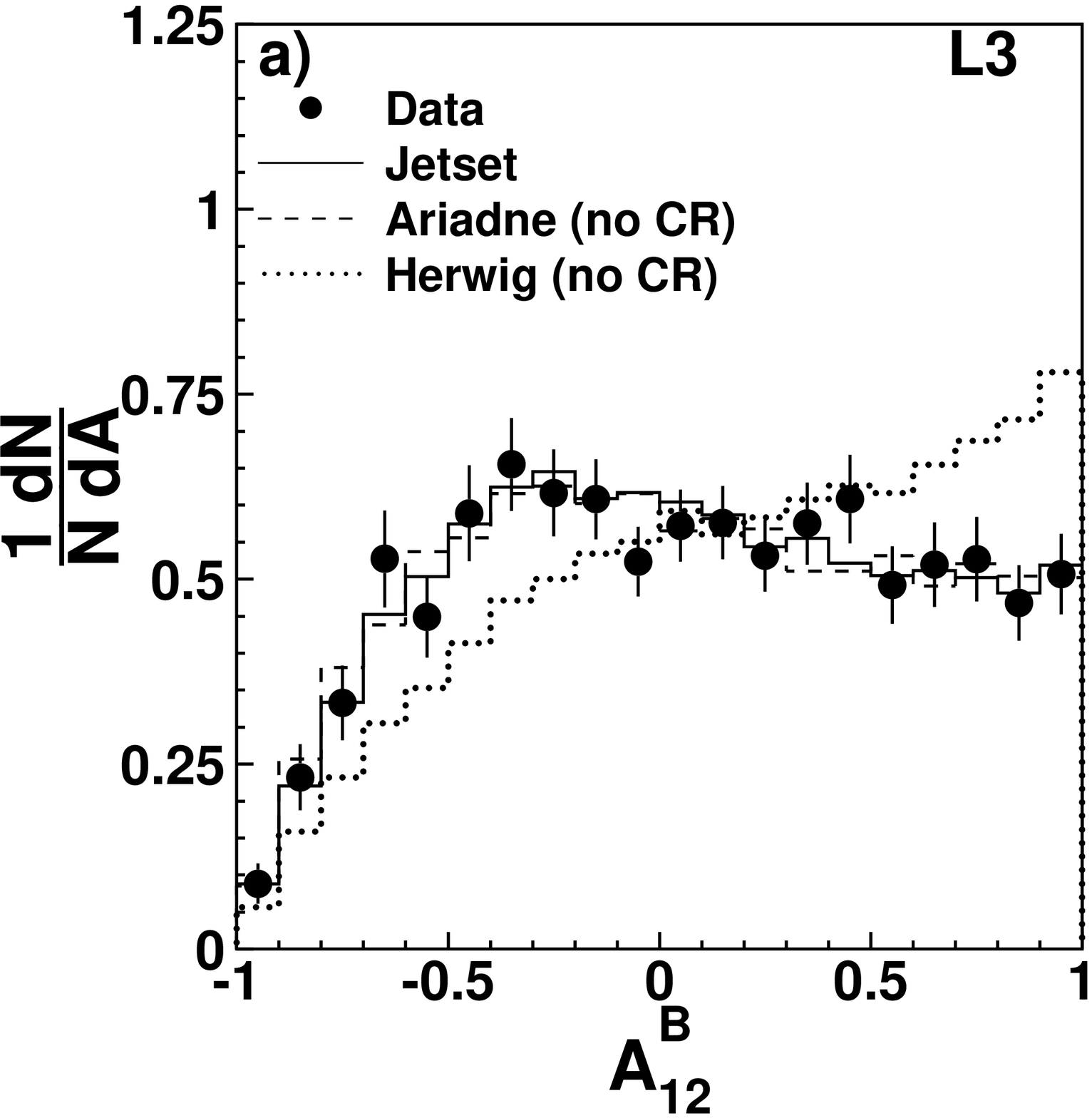}
\\ 
  \includegraphics[width=0.72\textwidth]{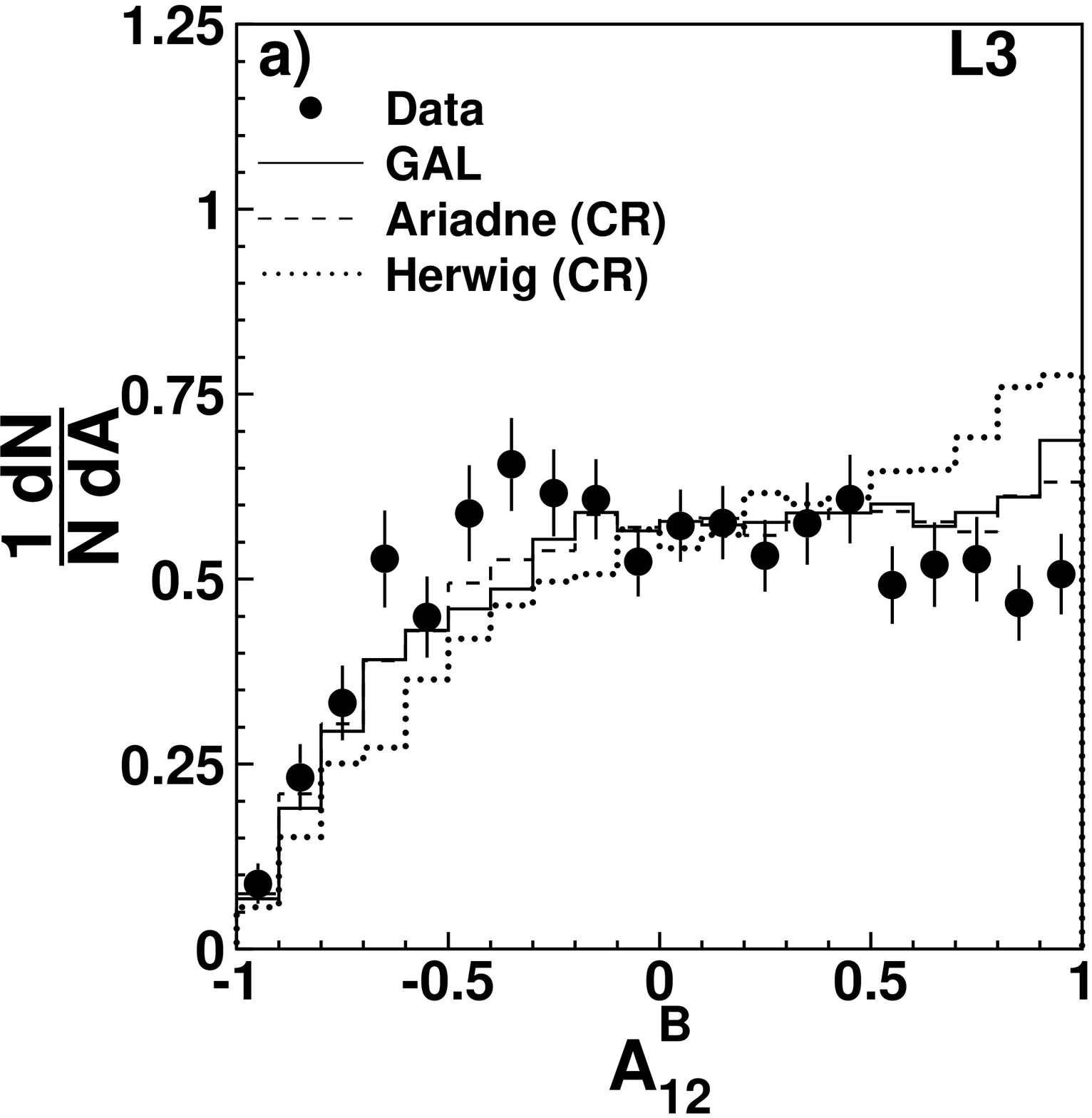}
\end{center}
\caption{A
         color-flow asymmetry variable for 3-jet events compared to MC predictions with and without
         color reconnection models.}
\label{fig:cras}
\end{figure}

The asymmetry distribution is shown in Fig.~\ref{fig:cras} and compared to the expectations of various
Monte Carlo models, both without and with a ``color reconnection'' (CR) algorithm.
Both {\sc jetset} and {\sc ariadne} without CR agree well with the data. However, when
the GAL model of Rathsman \cite{rathsman} is included in {\sc jetset} or when the CR model in
{\sc ariadne} is used, the models disagree with the data.  The agreement between {\sc herwig} and
the data is very poor both with and without its CR model.
The failure of the CR models here suggests that they are also inapplicable to the case of CR in
$\epem\rightarrow\mathrm{W}^+\mathrm{W}^-\rightarrow\mathrm{q}\bar{\mathrm{q}}\mathrm{q}\bar{\mathrm{q}}$.
 
\begin{thebibliography}{9}
 
  \bibitem{lepwg} LEP QCD Working Group,
  contributed paper, EPS HEP 2003, Aachen. 
%
%
%
%
%
%
%
%
 
  \bibitem{opalLow} OPAL Collab.,
      OPAL Physics Note PN519 (2003)
  \bibitem{aleph} ALEPH Collab.,
      ALEPH  2003-014.
 
 
 
  \bibitem{delphi19} DELPHI Collab.,
      DELPHI 2003-019 CONF 639.
  \bibitem{delphi26} DELPHI Collab.,
      DELPHI 2003-026 CONF 646.
  \bibitem{l3qcd} L3 Collab.,
          L3 note 2816 (2003).
 
  \bibitem{gary} J.W. Gary, talk in this session.
  \bibitem{rapgap} L3 Collab.,
    L3 note 2807 (2003).
  \bibitem{rathsman} J. Rathsman {\it et al.},  Phys. Lett.  {\bf B452} (1999) 364.
 
 
\end{thebibliography}
\end{document}